%% file: manuscript.tex
\documentclass[10pt,conference]{IEEEtran}

\usepackage{amsmath,amsfonts,amssymb}

\usepackage{tabularx}
\usepackage{multirow,multicol,makecell}
\usepackage{booktabs}
\usepackage{soul}
\usepackage{balance}

\usepackage{graphicx}
\usepackage{tikz}
\usetikzlibrary{arrows.meta,positioning,fit,calc,backgrounds}

\usepackage{framed}
\usepackage{enumitem}
\usepackage{float}
\usepackage{placeins}
\usepackage[table]{xcolor}

\usepackage{url}
\usepackage{hyperref}

\usepackage{caption}
\newcommand{\summarybox}[1]{%
{%
    \small
    \setlength{\OuterFrameSep}{0pt}%
    \definecolor{shadecolor}{HTML}{3A7F5A}%
    \def\FrameCommand{\colorbox{shadecolor}}%
    \MakeFramed{\hsize=\dimexpr\linewidth}%
      \noindent\textbf{\color{white} Takeaway.}%
    \endMakeFramed
    \nointerlineskip

    \definecolor{shadecolor}{HTML}{e2d1ab}
    \def\FrameCommand{\colorbox{shadecolor}}
    \MakeFramed{\hsize=\dimexpr\linewidth}
      \noindent{\color{black}#1}
    \endMakeFramed
}%
}

\AtBeginDocument{%
  }
\begin{document}
\IEEEoverridecommandlockouts

\title{Technical Debt Risks in AI-Assisted Software Development: Insights from Practitioners}

\author{
\IEEEauthorblockN{Nazia Jahan\IEEEauthorrefmark{1}, Nafisa Alam\IEEEauthorrefmark{1}, Sourov Debnath,\\ Arafat Rahman, S. M. Waesh Ahmed, Md Farhad Alam Bhuiyan}
\IEEEauthorblockA{
\textit{Penta Global Limited}\\
Dhaka, Bangladesh\\
\{nazia.jahan, nafisa.alam, sourov.debnath,\\ arafat, waesh.ahmed, farhad.alam\}@pentabd.com}
\and
\IEEEauthorblockN{Mia Mohammad Imran}
\IEEEauthorblockA{
\textit{Department of Computer Science}\\
\textit{Missouri University of Science and Technology}\\
Rolla, MO, USA\\
imranm@mst.edu}

\thanks{\IEEEauthorrefmark{1}These authors contributed equally to this work.}
}

% \author{
% \IEEEauthorblockN{First Author}
% \IEEEauthorblockA{
% \textit{Department of Computer Science}\\
% \textit{University Name}\\
% City, Country\\
% author1@university.edu}
% \and
% \IEEEauthorblockN{Second Author}
% \IEEEauthorblockA{
% \textit{Department of Computer Science}\\
% \textit{University Name}\\
% City, Country\\
% author2@university.edu}
% \and
% \IEEEauthorblockN{Third Author}
% \IEEEauthorblockA{
% \textit{Department of Computer Science}\\
% \textit{University Name}\\
% City, Country\\
% author3@university.edu}
% \and
% \IEEEauthorblockN{Fourth Author}
% \IEEEauthorblockA{
% \textit{Department of Computer Science}\\
% \textit{University Name}\\
% City, Country\\
% author4@university.edu}
% \and
% \IEEEauthorblockN{Fifth Author}
% \IEEEauthorblockA{
% \textit{Department of Computer Science}\\
% \textit{University Name}\\
% City, Country\\
% author5@university.edu}
% \and
% \IEEEauthorblockN{Sixth Author}
% \IEEEauthorblockA{
% \textit{Department of Computer Science}\\
% \textit{University Name}\\
% City, Country\\
% author6@university.edu}
% }

\maketitle

% \begingroup
% \renewcommand\thefootnote{*}
% \footnotetext{These authors contributed equally to this work.}
% \endgroup

\begin{abstract}

AI coding assistants increasingly support software development, raising concerns about technical debt associated with AI-generated code and development practices. We investigate these concerns through a sequential mixed-methods study involving 137 survey respondents and 20 follow-up interviews with software professionals in Bangladesh. Drawing on ten established technical debt dimensions, we examine perceived risks, contributing circumstances, maintenance consequences, validation practices, and developer understanding. Respondents expressed the strongest concerns about People Debt (73.0\%), Management Debt (70.1\%), and Architecture Debt (68.6\%), suggesting that risks extend beyond code quality. Participants associated ambiguous requirements, insufficient project context, and unrequested AI-generated changes with architectural inconsistencies, rework, and maintenance difficulties. Their experiences revealed contrasting outcomes: AI assistance supported refactoring, maintainability improvements, and testing, but also introduced rework and context misalignment. Although 89.8\% reported regularly verifying AI-generated code, participants expressed concerns about AI dependence, reduced code understanding, and review effort. They also emphasized the need for stronger architectural awareness, requirements clarification, and validation support in coding assistants. Our findings suggest that managing technical debt risks requires attention to project context, implementation decisions, and developer understanding alongside generated code quality.

\end{abstract}

\input{introduction}

\input{related}

\input{methodology}

\input{results}

\input{discussion}

\input{conclusion}

\balance 

\bibliographystyle{IEEEtran}
\bibliography{references}
\end{document}

%% file: introduction.tex
\section{Introduction}

Developers increasingly use generative artificial intelligence (GenAI) in software development \cite{liang2024usability, sergeyuk2025using} for code generation, debugging, testing, refactoring, and documentation \cite{sergeyuk2025using}. AI coding assistants help developers implement features, explore solutions, and modify existing systems. Barke et al.~\cite{barke2023grounded} identified exploration and acceleration as distinct modes of interaction with code-generating models, while Sergeyuk et al.~\cite{sergeyuk2025using} documented their use across development activities and highlighted concerns about reliability, project context, and developer control. As developers delegate increasingly complex tasks to AI assistants, these tools can generate substantial code changes with limited human intervention, raising concerns about additional review, integration, and long-term maintenance costs \cite{he2026speed}.

Technical debt (TD) refers to design or implementation constructs that are expedient in the short term but set up a technical context that can make future changes more costly or impossible~\cite{avgeriou2016managing}. Caglayan et al. proposed ten TD types: \textit{Architecture, Requirements, Configuration, Design, Documentation, Infrastructure, Management, People, Code, and Test debt}~\cite{caglayan2024revisiting}. AI-generated code can contain maintainability problems such as code smells, duplicated code, and missing documentation~\cite{siddiq2024quality,liu2026debt}. Such problems may become technical debt when they persist in the codebase and make later changes more costly~\cite{cunningham1992wycash,he2026speed,liu2026debt}. However, not every AI-generated code problem constitutes technical debt, as developers may revise or fix such problems before they create lasting maintenance liabilities~\cite{cunningham1992wycash, tabarsi2026reshaping, liu2026debt}. Investigating these risks therefore requires examining implementation problems alongside their development circumstances and consequences.

Technical debt risks also extend beyond individual code artifacts to the context in which developers use AI assistants. AI assistants may generate code that addresses the immediate prompt without the project-level context it needs, such as unstated requirements~\cite{liang2024usability, sergeyuk2025using}. Developers decide which tasks to delegate, how to specify their requirements, and how thoroughly to validate generated changes~\cite{sergeyuk2025using, barke2023grounded, tabarsi2026reshaping}. These decisions, combined with the speed and volume of AI-generated code, can affect how well developers understand the implementation, how much review effort it requires, and how maintainable it remains~\cite{vaithilingam2022expectation,storey2026technical,xu2025aiassisted}.

Prior studies have examined code-quality problems in AI-generated programs~\cite{liu2024refining}, TD in AI-enabled systems~\cite{recupito2024technical}, and TD risks in AI-assisted development and vibe coding~\cite{ehsani2026faster, niemeyer2026genai, he2026speed, mertala2026vibe, liu2026debt}. Repository studies primarily identify technical debt indicators in software artifacts, offering limited insight into the development decisions and circumstances behind them. Existing practitioner research also provides limited evidence connecting technical debt dimensions with developers' experiences and practices~\cite{recupito2024technical, mertala2026vibe}.

To address these gaps, we conduct a sequential mixed-methods study involving surveys and interviews with software development professionals in Bangladesh. We examine ten TD dimensions proposed by Caglayan et al.~\cite{caglayan2024revisiting} in relation to development decisions, maintenance consequences, mitigation practices, and developer understanding, guided by the following five research questions:

\begin{itemize}[leftmargin=*]
    \item \textbf{RQ1:} What types of technical debt do developers perceive as risks in AI-assisted software development?
    \item \textbf{RQ2:} What circumstances and development decisions contribute to problematic AI-generated implementations, and how can these introduce technical debt?
    \item \textbf{RQ3:} What consequences do developers experience from problematic AI-generated implementations, including those that create technical debt?
    \item \textbf{RQ4:} How do developers validate AI-generated code and manage or mitigate potential technical debt risks?
    \item \textbf{RQ5:} How does AI-assisted development influence developers' learning, work practices, understanding, and decision-making?
\end{itemize}

Our study comprises a survey of 137 software professionals, followed by interviews with 20 participants recruited from the survey respondents. We integrate quantitative survey results, thematic analysis of open-ended responses, and interview findings to examine perceived TD risks, contributing circumstances, maintenance consequences, and developer practices in AI-assisted software development.

Practitioners perceived TD risks across all ten dimensions, with People Debt (73.0\%) and Management Debt (70.1\%) receiving the highest agreement. Respondents identified ambiguous requirements and AI overreliance as prominent concerns, while qualitative accounts linked insufficient project context to rework and maintenance difficulties. Participants reported contrasting experiences, including improvements in refactoring, maintainability, and testing alongside rework, context misalignment, and maintenance challenges. Although most participants verified AI-generated code, manual review (78.1\%) remained more common than automated testing. Participants also reported learning benefits alongside concerns about AI dependence and reduced code understanding.

Our study makes three major contributions:

\begin{itemize}[leftmargin=*]
    \item \textbf{Socio-technical perspective:} We characterize practitioners' concerns about AI-associated technical debt across software quality, development processes, team coordination, and developer knowledge.

    \item \textbf{Pathways to technical debt:} We provide empirical evidence linking unclear requirements, insufficient project context, and unrequested AI-generated changes to architectural inconsistencies, rework, and maintenance difficulties.
    
    \item \textbf{Developers’ insights:} We identify practical challenges in delegating, reviewing, and maintaining AI-generated code, alongside developers' approaches to validation and preserving implementation understanding.
\end{itemize}

The survey and interview instruments, study data, annotation guidelines, and supporting materials are available at~\cite{replication2026}.

%% file: related.tex
\section{Related Work}

\subsection{Technical Debt in Software Engineering}
Research increasingly recognizes technical debt as a multidimensional, socio-technical phenomenon rather than a problem confined to source-code quality. Studies have established classifications of TD types, indicators, and management strategies~\cite{Li2015TechnicalDebt, alves2016identification, rios2018tertiary}. Caglayan et al. consolidated 120 indicators across ten dimensions~\cite{caglayan2024revisiting}. Their taxonomy provides the foundation for our study.

Empirical studies revealed how technical, organizational, and human factors contribute to technical debt. Ernst et al. identified architectural decisions as a major source of technical debt~\cite{ernst2015measure}, while Macia et al. demonstrated the limitations of code-level indicators in detecting architectural problems~\cite{macia2012anomalies}. Ahmad et al. highlighted the interplay between social, process, and people debts~\cite{Ahmad2022PeopleDebt}. Practitioner studies exposed organizational and process causes~\cite{rios2020practitioners}, shortcomings in management tools~\cite{biazotto2025understanding}, and gaps in decision traceability~\cite{carvalho2026traceability}. Recent work connected these challenges to stakeholder perspectives, team practices, and organizational structures, showing how technical debt emerges and persists through interactions between people and software systems~\cite{riquet2024debt, jorgensen2026experience}.

\subsection{AI-Assisted Software Development}
Research on AI coding assistants highlights productivity gains alongside challenges involving developer control and code quality. Cui et al. reported that developers with access to GitHub Copilot completed more tasks~\cite{cui2026effects}, while Barke et al. identified acceleration and exploration as distinct interaction modes~\cite{barke2023grounded}. However, Vaithilingam et al. observed difficulties understanding, modifying, and debugging generated code~\cite{vaithilingam2022expectation}. Broader surveys revealed concerns about trust, requirements compliance, insufficient project context, and the boundaries of task delegation~\cite{liang2024usability, sergeyuk2025using}. Mu et al. showed that asking clarifying questions before code generation improved the correctness of solutions for ambiguous requirements~\cite{mu2024clarifygpt}.

Verification and maintenance research exposed additional challenges. Tang et al. examined developers' validation and repair behaviors~\cite{tang2024validating}, while Tabarsi et al. documented practices involving code inspection, output checking, and testing~\cite{tabarsi2026reshaping}. Empirical evaluations identified correctness, maintainability, and security weaknesses in generated implementations~\cite{nguyen2022empirical, liu2024refining, pearce2022asleep, perry2023users}. Studies of downstream consequences reported contrasting findings: Xu et al.~\cite{xu2025aiassisted} associated AI adoption with increased rework and review effort, whereas Borg et al.~\cite{borg2026echoes} found no significant differences in subsequent maintenance performance. These results suggest that productivity gains, verification demands, and maintenance outcomes vary with development practices and contexts.

\subsection{Technical Debt in AI-Assisted Development}
Recent research examined technical debt risks associated with AI-assisted software development (Table~\ref{tab:related-td-studies}). Ehsani et al. identified code, design, and documentation debt alongside emerging concerns involving rapid integration and provenance~\cite{ehsani2026faster}. Moreschini et al. highlighted governance and maintainability challenges~\cite{moreschini2026evolution}. Gosu et al. found that existing approaches mainly addressed debt identification and repayment using prompt-based models, with limited support for broader technical debt management activities and human collaboration~\cite{gosu2026use}.

Empirical studies revealed persistent quality issues in AI-authored commits~\cite{liu2026debt}, while Niemeyer et al. reported different patterns of code, design, and architecture debt across project sizes~\cite{niemeyer2026genai}. Studies of self-admitted technical debt (SATD) identified postponed testing, incomplete adaptation, and requirements and design concerns~\cite{mujahid2026todo, brahmi2026characterizing}. These findings established evidence of maintenance risks in generated artifacts, while other researchers provided complementary evidence on technical debt risks in vibe coding~\cite{mertala2026vibe, fawzy2026vibe}. Storey's Triple Debt Model further connected code-level problems with developer understanding and implementation rationale~\cite{storey2026technical}, while Avgeriou et al. emphasized human responsibility and oversight in technical debt management~\cite{avgeriou2026technical}.

\noindent
\textbf{Our Research Positioning.}
Unlike studies that synthesized existing evidence~\cite{ehsani2026faster}, examined repository-level indicators~\cite{he2026speed, liu2026debt, niemeyer2026genai}, or focused on vibe coding~\cite{mertala2026vibe}, our study investigates technical-debt risks through software professionals' everyday experiences with AI-assisted development. Building on prior practitioner research~\cite{recupito2024technical, kara2026hierarchical, mertala2026vibe}, we examine ten established technical-debt dimensions alongside contributing development circumstances, maintenance consequences, mitigation practices, and developer understanding.

\begin{table}[tb]
\centering
\scriptsize
\caption{Relevant AI-related technical debt studies (2024--2026).}
\label{tab:related-td-studies}
\resizebox{\columnwidth}{!}
{
\begin{tabular}{|p{2.2cm}|p{2.2cm}|p{5cm}|}
\hline
\textbf{Study} & \textbf{Data Source} & \textbf{Research Focus} \\
\hline
Ehsani et al.~\cite{ehsani2026faster} & 104 studies (includes gray literature) & Existing TD types, proposed new TD types, TD causes, and mitigation in AI-assisted development \\
Moreschini et al.~\cite{moreschini2026evolution} & 61 studies & TD types and challenges from DevOps to GenAI \\
Gosu et al.~\cite{gosu2026use} & 29 studies & GenAI for identifying and repaying TD \\
\hline
Liu et al.$^\dagger$~\cite{liu2026debt} & 302.6k commits& Persistence of issues in AI-authored commits \\
Niemeyer et al.~\cite{niemeyer2026genai} & 1,091 repositories & Code, design, and architecture debt after GenAI adoption \\
Mujahid et al.~\cite{mujahid2026todo} & 6,540 code comments & SATD~\cite{potdar2014exploratory} associated with AI-generated code \\
Brahmi et al.~\cite{brahmi2026characterizing} & 525 code comments & SATD generated by AI coding agents \\
Xu et al.$^\dagger$~\cite{xu2025aiassisted} & 2,755 repositories & Productivity, rework, and maintenance burden \\
\hline
Bakke et al.~\cite{bakke2026reducing} & 23 teams & GenAI-assisted refactoring to reduce TD \\
Recupito et al.~\cite{recupito2024technical} & Survey (53) & Code and architecture debt in AI-enabled systems \\
Kara et al.$^\dagger$~\cite{kara2026hierarchical} & Interview (18) & Proposed ten TD categories in GenAI-enabled systems \\
Mertala \& Ivanisevic$\ddagger$~\cite{mertala2026vibe} & Survey (61); interview (7) & TD risks in vibe coding \\
\hline
\textbf{Our Study} & \textbf{Survey (137); interview (20)} & \textbf{Ten TD dimensions, causes, consequences, and practices in AI-assisted development among professionals} \\
\hline
\end{tabular}
}
\par\smallskip\noindent
\parbox{\columnwidth}{\scriptsize\textit{Note:} $\dagger$Preprint; $\ddagger$Thesis}
\end{table}

%% file: methodology.tex
\section{Methodology}
We adopt a sequential mixed-methods design~\cite{maxwell2003mixed, ivankova2006using}, beginning with an online survey followed by semi-structured interviews. We analyze closed-ended survey responses using descriptive statistics and apply thematic analysis~\cite{braun2006using} to open-ended responses. We use the survey findings to develop the interview protocol and recruit participants for follow-up interviews. Finally, we analyze interview accounts to further explore the experiences, implementation decisions, and development practices identified through the survey.

\subsection{Survey Design}
We develop a 35-question survey and distribute through Google Forms. The survey comprised 32 closed-ended and three open-ended questions across five areas: participant background (Q1-Q4), AI-tool usage (Q5-Q8), attitudes and practices (Q9-Q12), technical debt risks and experiences (Q13-Q32), and open-ended experiences (Q33-Q35). It took approximately 25-30 minutes to complete the survey. Table~\ref{tab:survey-design} presents the survey questions and response formats.

\noindent
\textul{\em Participant Background and AI-Tool Usage.}
Questions Q1-Q4 collected participants' gender, professional roles, software development experience, and frequently used programming languages. Questions Q5-Q8 examined whether participants used AI tools for software development, which tools they used, how frequently they used them, and their purposes. Participants selected one response for gender, role, experience, AI use, and usage frequency. They could select multiple programming languages, AI tools, and usage purposes.

\input{tables/survey}

\noindent
\textul{\em Attitudes and Development Practices.}
Questions Q9-Q12 examined perceived helpfulness, trust in AI-generated code without modification, verification frequency, and learning benefits. Q9 offered five categories ranging from Very helpful to Not helpful at all, while Q10 offered five categories ranging from Completely trust to No trust at all. Q11 measured verification frequency using Always, Often, Sometimes, Rarely, and Never. Q12 measured learning benefits using Significantly, Somewhat, Rarely, and Never. We treated the questions as ordered response in Likert scales.

\noindent
\textul{\em Technical Debt Related Questions.}
Q13-Q32 examine ten technical debt dimensions from the taxonomy of Caglayan and Ozcan-Top~\cite{caglayan2024revisiting}: Architecture, Code, Configuration, Design, Documentation, Infrastructure, Management, People, Requirements, and Test debt. For each dimension, we pair a five-point agreement statement with a multiple-selection follow-up question. We develop the follow-up options by drawing on technical debt indicators and classifications~\cite{caglayan2024revisiting, rios2018tertiary, alves2016identification, melo2022identification, maldonado2015detecting, suryanarayana2014refactoring} and adapting them to AI-assisted development~\cite{ehsani2026faster}. This approach allows us to cover established technical debt problems while incorporating AI-specific concerns, including overreliance on generated solutions, dependence on AI tools, and reduced understanding of generated code. The agreement statements capture participants' perceptions of potential technical debt risks, while the follow-up questions examine issues they experience or observe, including duplicated code, architectural inconsistencies, dependency conflicts, and requirement mismatches, as well as their practices for validating AI-generated code. Participants can select multiple predefined options, choose ``None'' if none apply, or select ``Other'' to provide a free-text response.

\noindent
\textul{\em Open-Ended Questions.}
We ask the participants to describe their experiences beyond the predefined response options in three open-ended questions. Q33 explored situations in which AI tools helped reduce or manage technical debt, Q34 examined problems participants encountered during AI-assisted development, and Q35 collected suggestions for improving AI coding assistants. These questions allowed participants to describe implementation circumstances, consequences, and potential improvements in their own words.

\subsection{Survey Participants}

\noindent
\textul{\em Recruitment and Eligibility.}
We recruited software development professionals working in Bangladesh through voluntary, non-probability sampling. We distributed the Google Forms survey through social media and professional networks, including LinkedIn, Facebook, and Discord, using public posts and direct sharing. The survey was conducted from June 6 to June 16, 2026. Respondents provided informed consent before completing the questionnaire. We received 137 responses. We refer the survey participants as S1-S137. 

\noindent
\textul{\em Demographic and Professional Background.}
Table~\ref{tab:survey-participants} summarizes participants' gender, software development experience, and professional roles. Practitioners with three to five years of experience constituted the largest group (40.1\%). Backend developers (28.5\%) and full-stack developers (21.9\%) represented the most common roles. We grouped less frequent roles under ``Other roles.''

\begin{table}[!tb]
\scriptsize
\centering
\caption{Survey participant characteristics ($N=137$).}
\label{tab:survey-participants}
{
\begin{tabular}{|l|c|l|c|}
\hline
\multicolumn{4}{|l|}{\textbf{AI-tool use for software development}} \\
\hline
\multicolumn{2}{|l|}{Yes (included in analysis)} & \multicolumn{2}{l|}{137 (100\% of 137)} \\
\hline
\multicolumn{4}{|l|}{\textbf{Gender:} Male: 129 (94.2\%); Female: 8 (5.8\%)} \\
\hline
\textbf{Experience} & \textbf{n (\%)} & \textbf{Role} & \textbf{n (\%)} \\
\hline
Less than one year & 10 (7.3\%) & Backend developer & 39 (28.5\%) \\
One to two years & 28 (20.4\%) & Full-stack developer & 30 (21.9\%) \\
Three to five years & 55 (40.1\%) & Data/ML engineer & 20 (14.6\%) \\
Six to ten years & 32 (23.4\%) & QA/Test engineer & 11 (8.0\%) \\
More than ten years & 12 (8.8\%) & Frontend developer & 10 (7.3\%) \\
 & & Other roles & 27 (19.7\%) \\
\hline
\end{tabular}
}
\end{table}

\subsection{Interview Design}
We develop an eight-question semi-structured interview protocol (Table~\ref{tab:interview-questions}) to explore the issues identified in our survey analysis in greater depth. The protocol covers AI-tool usage and task delegation (I-Q1--I-Q2), changes beyond the requested scope (I-Q3), handling of ambiguous requirements (I-Q4), implementation intent (I-Q5), learning with unfamiliar technologies (I-Q6), validation practices (I-Q7), and overlooked risks (I-Q8). We ask the core questions to all interviewees and use follow-up questions to explore their experiences in greater depth. The wording of some questions varies with the flow of each conversation.

\input{tables/interview}

\noindent
\textul{\em Participant Recruitment.}
We invite interview participants from the survey respondents based on their reported experiences and willingness to participate, while seeking diversity in professional roles and software development experience. Of the respondents who expressed willingness to participate in a follow-up interview, we recruit 20 participants for the interview study  (Table~\ref{tab:participants}). We obtain separate consent for interview participation and audio recording. We identify interview participants as P1-P20 throughout the study. 

\noindent
\textul{\em Interview Conduct.}
Five authors conduct 20 interviews between September 1 and 10, 2026, either in person or remotely via Zoom and Google Meet. The interviews last approximately 10-30 minutes and take place in Bangla or English, depending on participants' preferences. With participants' consent, we record the interviews for subsequent transcription and qualitative analysis. We generate initial transcripts manually and with YouTube's automatic captions for private uploads~\cite{youtube_captions}. We then translate the transcripts into English using Claude Opus 5.5~\cite{anthropic_opus55}. In each step, tw authors independently check the transcripts and English translations against the original recordings and correct any inaccuracies.

\input{tables/open-end-themes}

\subsection{Quantitative Data Analysis}
We use descriptive statistics to analyze the 137 valid survey responses. For each question, we calculate response frequencies and percentages using the number of valid responses as the denominator. We report Likert-scale distributions across their original response categories and exclude missing or invalid answers only from the corresponding question. For multiple-selection questions, we count the respondents who select each option and calculate the corresponding percentages. Because respondents can select more than one option, the percentages for these questions can exceed 100\% in total.

\subsection{Qualitative Data Analysis}

\paragraph{Open-Ended Survey Responses}
We conduct an inductive thematic analysis~\cite{braun2006using} of responses to Q33-Q35. We exclude blank and non-substantive responses, retaining 118 responses for Q33, 109 for Q34, and 112 for Q35.

\noindent
\textul{\em Annotation Procedure.}
Two annotators (authors of this paper) independently examine the responses and assign initial codes to recurring concepts and experiences. They compare their codes, discuss differences, and group related codes into broader themes. Through eight discussion sessions, they refine theme definitions, resolve disagreements, and distinguish overlapping categories. Each annotator devotes approximately 120 person-hours to this process. We allow multiple codes per response when participants describe distinct experiences or concerns, so the combined theme frequencies may exceed the number of valid responses.

\noindent
\textul{\em Question-Specific Analysis.}
We develop themes independently for Q33, Q34, and Q35, which address technical debt reduction and software-quality improvements (Table~\ref{tab:q33-themes}), implementation problems and maintenance consequences (Table~\ref{tab:q34-themes}), and desired improvements to AI coding assistants (Table~\ref{tab:q35-themes}), respectively.

\paragraph{Interview Data}
Two authors review interview responses by question and compare participants' experiences and practices. They identify recurring issues, group similar accounts, and record how many interviewees report each issue. We count each participant once per category and calculate percentages based on the number of respondents to the corresponding question. We examine these patterns alongside individual accounts to understand implementation problems, maintenance consequences, and developers' validation and decision-making practices.

\subsection{Ethical Considerations}
We obtained informed consent for survey participation and separate consent for interviews and audio recording. Participation was voluntary, and participants could withdraw. We anonymize all responses to protect participants' identities and maintain confidentiality.

%% file: tables/survey.tex
\begin{table}[tb]
\centering
\scriptsize
\caption{Survey questions and response types.}
\label{tab:survey-design}
\resizebox{\columnwidth}{!}
{
\begin{tabular}{|l|p{8cm}|c|}
\hline
\textbf{ID} & \textbf{Question or statement} & \textbf{Type} \\
\hline
\multicolumn{3}{|l|}{\textbf{AI-tool usage}} \\
\hline
Q5 & Do you use AI tools for software development? & YN$\dagger$ \\
Q6 & Which AI tools or agents do you use regularly? & MS$\dagger$ \\
Q7 & How often do you use AI tools while developing software? & SS$\dagger$ \\
Q8 & For what purposes do you use AI tools? & MS \\
\hline
\multicolumn{3}{|l|}{\textbf{Attitudes and practices}} \\
\hline
Q9 & How helpful are AI tools in your daily development work? & L5$\dagger$ \\
Q10 & How much do you trust AI-generated code without modification? & L5 \\
Q11 & How often do you verify AI-generated code before using it? & L5 \\
Q12 & Have AI tools helped you learn new concepts or technologies? & L4$\dagger$ \\
\hline
\multicolumn{3}{|l|}{\textbf{Architecture debt}} \\
\hline
Q13 & AI-generated solutions may negatively affect long-term system architecture. & L5 \\
Q14 & Which architecture-related issues have you experienced from AI-generated code? & MS \\
\hline
\multicolumn{3}{|l|}{\textbf{Code debt}} \\
\hline
Q15 & AI-generated code often increases difficult-to-maintain or low-quality code. & L5 \\
Q16 & Which code-related problems have you experienced from AI-generated code? & MS \\
\hline
\multicolumn{3}{|l|}{\textbf{Configuration debt}} \\
\hline
Q17 & AI-generated solutions may introduce configuration or dependency management issues. & L5 \\
Q18 & Which configuration-related issues have you noticed? & MS \\
\hline
\multicolumn{3}{|l|}{\textbf{Design debt}} \\
\hline
Q19 & AI tools sometimes suggest solutions that conflict with good software design principles. & L5 \\
Q20 & Which design-related issues have you noticed in AI-generated code? & MS \\
\hline
\multicolumn{3}{|l|}{\textbf{Documentation debt}} \\
\hline
Q21 & AI-assisted development may lead to incomplete or low-quality documentation. & L5 \\
Q22 & What documentation problems have you experienced with AI-generated content? & MS \\
\hline
\multicolumn{3}{|l|}{\textbf{Infrastructure debt}} \\
\hline
Q23 & AI-generated solutions may introduce infrastructure or operational inefficiencies. & L5 \\
Q24 & Which infrastructure-related problems have you experienced? & MS \\
\hline
\multicolumn{3}{|l|}{\textbf{Management debt}} \\
\hline
Q25 & Heavy reliance on AI tools may create management or workflow-related challenges within software teams. & L5 \\
Q26 & Which management-related issues have you observed due to AI-assisted development? & MS \\
\hline
\multicolumn{3}{|l|}{\textbf{People debt}} \\
\hline
Q27 & AI-assisted development may negatively affect collaboration, learning, or developer skill growth. & L5 \\
Q28 & Which people-related challenges have you experienced due to AI-generated code? & MS \\
\hline
\multicolumn{3}{|l|}{\textbf{Requirements debt}} \\
\hline
Q29 & AI-generated solutions may fail to fully reflect actual business requirements. & L5 \\
Q30 & In which situations do AI tools fail most regarding requirements? & MS \\
\hline
\multicolumn{3}{|l|}{\textbf{Test debt}} \\
\hline
Q31 & Developers may rely on AI-generated code without proper testing. & L5 \\
Q32 & How do you usually validate AI-generated code before production use? & MS \\
\hline
\multicolumn{3}{|l|}{\textbf{Open-ended questions}} \\
\hline
Q33 & Describe a situation where AI tools helped reduce technical debt and improved software quality in your development work. & OE$\dagger$ \\
Q34 & Describe a situation where AI-generated code or AI-assisted development introduced technical debt or caused problems in your software project. & OE \\
Q35 & What improvements would you like to see in AI coding assistants to reduce the risk of AI-generated technical debt? & OE \\
\hline
\end{tabular}
}
\par\smallskip\noindent
\parbox{\columnwidth}{\scriptsize\textit{$\dagger$Note:}  YN = Yes/No; SS = single selection; MS = multiple selection; L5 = five-point Likert scale; L4 = four-point Likert scale; OE = open-ended.}
\end{table}

%% file: tables/interview.tex
\begin{table}[tb]
\centering
\scriptsize
\caption{Semi-structured interview questions.}
\label{tab:interview-questions}
\setlength{\tabcolsep}{3pt}
\resizebox{\columnwidth}{!}
{
\begin{tabular}{|p{0.5cm}|p{9cm}|}
\hline
 & \textbf{Question} \\
\hline
I-Q1 & Which AI coding tools do you currently use? Free or paid versions? Do you primarily use conversational tools for specific questions or code snippets, or agentic tools that can access and modify larger portions of your codebase? \\
\hline
I-Q2 & At what point in a task do you decide to hand over control to an AI agent instead of doing the work yourself? What usually influences that decision? \\
\hline
I-Q3 & Can you describe a scenario where an AI agent made changes beyond your instructions? What additional changes did it make, and how did you handle them? \\
\hline
I-Q4 & How does AI typically respond to ambiguous requirements, business rules, or edge cases? Does it seek clarification or proceed based on assumptions? \\
\hline
I-Q5 & When asking an AI agent to implement a task, how do you make sure it understands what you want and why? Can you remember a situation where you or another developer later had to change AI-generated code but did not know the original reason behind the implementation? What happened? \\
\hline
I-Q6 & When working with unfamiliar concepts or technologies, how do you use AI? Do you first seek explanations before implementation, or delegate the implementation directly? What influences your approach? \\
\hline
I-Q7 & Before accepting AI-generated changes into your project, how do you decide that the code is ``good enough'' to keep? When you accept a suggestion or solution from an AI agent, do you usually verify that it is correct before using it? If so, how do you verify it? \\ 
\hline
I-Q8 & Based on your experience, what is the most significant overlooked risk or problem associated with AI-assisted software development? \\
\hline
\end{tabular}
}
\end{table}

\begin{table}[!tb]
\scriptsize
\centering
\caption{Demographics of the interview participants.}
\label{tab:participants}
\resizebox{\columnwidth}{!}
{
\begin{tabular}{|l|l|c|l|l|c|}
\hline
\multicolumn{6}{|p{10cm}|}{\textbf{Gender:} Male: 19 (95\%); Female: 1 (5\%)} \\
\hline
& \textbf{Participant role} & \textbf{Experience} 
& & \textbf{Participant role} & \textbf{Experience} \\
\hline
P1  & Data/ML Engineer & 3--5 yrs  & P2  & QA/Test Engineer & 3--5 yrs \\  
P3  & Full-stack Developer & 3--5 yrs  & P4  & Backend Developer & 1--2 yrs \\  
P5  & QA/Test Engineer & 3--5 yrs  & P6  & Backend Developer & 10+ yrs \\  
P7  & Full-stack Developer & 10+ yrs & P8  & QA/Test Engineer & 3--5 yrs \\ 
P9  & Frontend Developer & 3--5 yrs  & P10 & Data/ML Engineer & 3--5 yrs \\ 
P11 & Frontend Developer & 3--5 yrs  & P12 & Data/ML Engineer & 1--2 yrs \\  
P13 & Full-stack Developer & 1--2 yrs  & P14 & Mobile App Developer & 3--5 yrs \\ 
P15 & Frontend Developer & 3--5 yrs  & P16 & Frontend Developer & 3--5 yrs \\  
P17 & Backend Developer & 6--10 yrs & P18 & Backend Developer & 10+ yrs \\ 
P19 & Frontend Developer & 1--2 yrs  & P20 & Backend Developer & 6--10 yrs \\ \hline
\end{tabular}
}
\end{table}

%% file: tables/open-end-themes.tex
\begin{table}[tb]
\centering
\scriptsize
\caption{Q33 themes: Reported AI-assisted technical debt reduction and software quality improvements. Responses may contain multiple themes.}
\label{tab:q33-themes}
\resizebox{\columnwidth}{!}
{
\begin{tabular}{|p{2.2cm}|p{6cm}|c|} \hline
\textbf{Theme} & \textbf{Definition} & \textbf{n} \\ \hline
Refactoring & Restructuring existing code without changing its behavior. & 46 \\ \hline
Productivity & Increasing development output while reducing time and effort. & 37 \\ \hline
Code Quality \& Maintainability & Improving code quality attributes, including readability, consistency, reliability, and ease of maintenance. & 36 \\ \hline
Test Generation & Generating test cases/test scripts to verify software behavior. & 31 \\ \hline
Code Generation & Generating source code, boilerplate, or scripts from specifications or prompts. & 31 \\ \hline
Code Comprehension & Understanding the structure, behavior, and logic of the code. & 25 \\ \hline
Code Review & Examining source code to identify defects and recommend improvements. & 18 \\ \hline
Architecture Design & Defining system structure, components, interfaces, and their interactions. & 18 \\ \hline
Debugging & Identifying, diagnosing, and correcting software defects. & 18 \\ \hline
Documentation & Creating and updating documentation and code comments. & 17 \\ \hline
Legacy System Maintenance & Modifying and maintaining existing legacy software systems. & 15 \\ \hline
Planning & Defining requirements, decomposing tasks, and organizing development activities. & 14 \\ \hline
Prototyping & Developing preliminary implementations to evaluate ideas/feasibility. & 6 \\ \hline
Performance Optimization & Improving execution speed, resource utilization, and computational efficiency. & 6 \\ \hline
Code Migration & Adapting software to different languages, frameworks, or platform versions. & 5 \\ \hline

Miscellaneous & Generic responses not covered by other themes. & 8 \\ \hline

\end{tabular}
}
\end{table}

\begin{table}[tb]
\centering
\scriptsize
\caption{Q34 themes: Reported technical debt and problems introduced by AI-assisted development. Responses may contain multiple themes.}
\label{tab:q34-themes}
\resizebox{\columnwidth}{!}{
\begin{tabular}{|p{2.2cm}|p{6cm}|c|} \hline
\textbf{Theme} & \textbf{Definition} & \textbf{n} \\
\hline
Rework & Additional effort to rewrite, refactor, correct, or revert AI-generated code. & 33 \\
\hline
Maintainability & Code that is difficult to understand, modify, or maintain. & 26 \\
\hline
Context Misalignment & Generated code that overlooks project requirements, constraints, or context. & 23 \\
\hline
Architectural Inconsistency & Violations of established architecture, design conventions, or coding standards. & 22 \\
\hline
Overengineering & Introduction of unnecessary complexity, abstractions, components, or functionality. & 21 \\
\hline
Incorrect or Hallucinated Output & Incorrect, fabricated, or irrelevant AI-generated code or information. & 20 \\
\hline
Software Reliability & New bugs, regressions, or runtime failures caused by AI-generated code. & 18 \\
\hline
AI Overreliance & Excessive AI reliance leading to cognitive dependence and weakened developer knowledge and skills. & 16 \\ \hline
Code Duplication & Repeated code or logic across functions, files, or modules. & 15 \\
\hline
Review Overhead & Extra effort to review and validate AI-generated code. & 11 \\
\hline
Performance & Inefficient code causing slow execution, excessive resource use, or performance bottlenecks. & 9 \\
\hline
Testing & Missing, inadequate, or unreliable tests, or AI actions that disrupt test suites. & 8 \\
\hline
Debugging Difficulties & Challenges in debugging AI-generated code due to limited context and added complexity. & 7 \\ \hline
AI Tool Limitations & Constraints and shortcomings of specific AI tools, models, and versions. & 5 \\ \hline
Domain-Specific Knowledge Gaps & Limited understanding of specialized domains, technologies, and their requirements. & 5 \\
\hline
Security & Vulnerable code or suggestions of unsafe practices. & 4 \\
\hline
Dependency Management & Version incompatibilities, dependency conflicts, or unnecessary dependencies. & 4 \\
\hline
Team Coordination & Inconsistent AI-use practices among the team members. & 4 \\
\hline
Infrastructure & Incorrect, broken, or oversized infrastructure configurations. & 2 \\
\hline
Legacy System Migration & Incorrect transformations during legacy migrations or upgrades. & 2 \\
\hline
Miscellaneous & Generic responses not covered by other themes. & 14 \\
\hline
\end{tabular}
}
\end{table}

\begin{table}[tb]
\centering
\scriptsize
\caption{Q35 themes: Suggested improvements to AI coding assistants for reducing technical debt risks. Responses may contain multiple themes.}
\label{tab:q35-themes}
\resizebox{\columnwidth}{!}{
\begin{tabular}{|p{2.2cm}|p{6cm}|c|} \hline
\textbf{Theme} & \textbf{Definition} & \textbf{n} \\
\hline
Architectural Awareness & Adhering to existing system architecture, design patterns, and coding conventions. & 30 \\
\hline
Risk Identification & Identifying and warning about uncertain, risky, or potentially problematic existing/generated code. & 18 \\
\hline
Coding Best Practices & Following established coding standards and principles for readable, modular, and maintainable code. & 17 \\
\hline
Testing Support & Generating tests and identifying edge cases to verify code correctness. & 17 \\
\hline
Requirements Understanding & Correctly interpreting requirements, business rules, and developer intent. & 16 \\
\hline
Output Reliability & Ensuring accurate, consistent, and contextually relevant outputs while minimizing hallucinations. & 13 \\\hline
Explainability & Explaining generated code, design decisions, and trade-offs. & 13 \\
\hline
Clarification Requests & Seeking clarification when requirements or instructions are ambiguous or incomplete. & 13 \\
\hline
Code Reuse & Reusing existing components and preventing code duplication. & 13 \\
\hline
Automated Code Validation & Automatically checking generated code for correctness and quality before acceptance. & 12 \\
\hline
Developer Skill Development & Enhancing developer skills through learning, training, and effective use of AI tools. & 11 \\ \hline
Security Validation & Identifying security vulnerabilities in the generated code. & 11 \\ \hline
Human Oversight & Maintaining developer control through review, validation, and guidance of AI-generated code. & 9 \\
\hline
Structured Planning & Outlining implementation steps and specifications before code generation. & 9 \\
\hline
Debt Detection & Identifying technical debt, unnecessary complexity, and maintainability concerns. & 8 \\
\hline
Context Management & Maintaining, retrieving, and effectively utilizing relevant project context to ensure continuity and consistency across tasks and interactions. & 8 \\
\hline
Change Scope Control & Limiting modifications to the requested scope while preserving existing functionality. & 6 \\
\hline
Documentation & Generating and maintaining documentation for code changes. & 5 \\
\hline
Cost \& Accessibility & Minimizing token-related constraints and computational costs while improving the affordability and accessibility of AI tools. & 3 \\
\hline
Miscellaneous & Generic responses not covered by other themes. & 3 \\
\hline
\end{tabular}
}
\end{table}

%% file: results.tex
\section{Results}

\subsection{RQ1: What types of technical debt do developers perceive as risks in AI-assisted software development?}

\input{results/rq1}

\subsection{RQ2: What circumstances and development decisions contribute to problematic AI-generated implementations, and how can these introduce technical debt?}

\input{results/rq2}

\subsection{RQ3: What consequences do practitioners experience from problematic AI-generated implementations, including those that create technical debt?}

\input{results/rq3}

\subsection{RQ4: How do developers validate AI-generated code and manage or mitigate potential technical debt risks?}

\input{results/rq4}

\subsection{RQ5: How does AI-assisted development influence developers' learning, work practices, understanding, and decision-making?}

\input{results/rq5}

%% file: results/rq1.tex
To answer RQ1, we examine practitioners' perceptions of ten established technical debt dimensions alongside their experiences from open-ended survey responses and interviews. Table~\ref{tab:rq1_likert} shows that respondents perceived technical debt risks across all ten dimensions, with the strongest concerns involving \textit{People Debt} (73.0\%), \textit{Management Debt} (70.1\%), and \textit{Architecture Debt} (68.6\%). In contrast, \textit{Documentation Debt} (31.4\%), \textit{Test Debt} (20.4\%), and \textit{Requirements Debt} (16.1\%) received the highest combined disagreement. \textit{Infrastructure Debt} had the highest proportion of neutral responses (48.2\%).

The open-ended survey responses provided examples of implementation and maintenance problems associated with these dimensions. In Q34 (Table~\ref{tab:q34-themes}), 109 respondents provided valid accounts. The most frequent themes included Rework, Maintainability, and Context Misalignment, which we examine in RQ3 and RQ2, followed by Architectural Inconsistency (22, 20.2\%), Overengineering (21, 19.3\%), Incorrect or Hallucinated Output (20, 18.3\%), and Software Reliability (18, 16.5\%). S10 reported that AI introduced unnecessary components that required removal in a later release, while S107 described an AI-assisted refactoring that broke an existing business requirement. Respondents also reported Testing (8, 7.3\%), Dependency Management (4, 3.7\%), and Security (4, 3.7\%) issues. For example, S63 described flaky tests that produced false-positive results, while S94 reported that an assistant deleted a failing unit-test suite.

Interview participants described similar difficulties involving implementation scope, requirements, and code understanding. All 20 interviewees reported that AI made changes beyond their requests. P10, for example, requested a simple function but received an entire class with unnecessary file operations, so P10 discarded the solution and reimplemented it. When requirements lacked clarity (I-Q4), 11 interviewees reported that AI sometimes proceeded on assumptions rather than seeking clarification. P12 explained, \emph{``if I don't explicitly mention this, it generally starts making assumptions and proceeds with the implementation.''} In addition, 16 interviewees (80.0\%) reported difficulty recovering the rationale behind AI-generated code (I-Q5), and 7 of them attributed this difficulty to missing comments or documentation. P20 connected this lack of understanding to maintainability: \emph{``It writes code its own way, so you can't understand the code's complexity: what it wrote and how.''}

People and management-related concerns also emerged from both sources. In Q34, 16 respondents (14.7\%) described AI Overreliance, while 11 (10.1\%) identified Review Overhead. S84 described difficulties modifying and debugging code they had accepted without fully understanding, while S118 reported that after AI introduced unrequested functions, \emph{``we lost the implementation knowledge.''} Similarly, 11 of 20 interviewees (55.0\%) named blind trust, overreliance, or reduced understanding, learning, or code ownership as the biggest overlooked risk (I-Q8). P4 pointed to \emph{``Reliability, blind faith, and the tendency of tools to agree with anything you say.''}

\summarybox{
Respondents perceived technical debt risks across all ten dimensions, with the strongest agreement for People Debt (73.0\%), Management Debt (70.1\%), and Architecture Debt (68.6\%). Their accounts highlighted rework, architectural inconsistencies, maintenance difficulties, and reduced understanding of AI-generated code.}

\begin{table}[tb]
\centering
\scriptsize

\caption{Responses of technical debt dimensions in AI-assisted software development ($N=137$). }
\label{tab:rq1_likert}
% \resizebox{\columnwidth}{!}
{
\begin{tabular}{|l|l|c|c|c|c|c|} \hline
\textbf{Q} & \textbf{TD Dimension} & \textbf{SD}$\dagger$ & \textbf{D}$\dagger$ & \textbf{N}$\dagger$ & \textbf{A}$\dagger$ & \textbf{SA}$\dagger$ \\
\hline
Q13 & Architecture Debt & 1.5\% & 8.0\% & 21.9\% & 43.1\% & 25.5\% \\
Q15 & Code Debt & 2.2\% & 9.5\% & 23.4\% & 41.6\% & 23.4\% \\
Q17 & Configuration Debt & 0.7\% & 7.3\% & 29.9\% & 47.4\% & 14.6\% \\
Q19 & Design Debt & 1.5\% & 7.3\% & 27.0\% & 48.9\% & 15.3\% \\
Q21 & Documentation Debt & 5.8\% & 25.5\% & 29.2\% & 24.8\% & 14.6\% \\
Q23 & Infrastructure Debt & 0.7\% & 6.6\% & 48.2\% & 34.3\% & 10.2\% \\
Q25 & Management Debt & 1.5\% & 5.1\% & 23.4\% & 43.8\% & 26.3\% \\
Q27 & People Debt & 0.7\% & 5.1\% & 21.2\% & 38.0\% & 35.0\% \\
Q29 & Requirements Debt & 0.7\% & 15.3\% & 25.5\% & 42.3\% & 16.1\% \\
Q31 & Test Debt & 13.1\% & 7.3\% & 16.1\% & 38.7\% & 24.8\% \\
\hline
\end{tabular}
}
\par\smallskip\noindent
\parbox{\columnwidth}{\scriptsize\textit{$\dagger$Note:} SD = Strongly Disagree, D = Disagree, N = Neutral, A = Agree, and SA = Strongly Agree.}
\end{table}

%% file: results/rq2.tex
To answer RQ2, we examine the development circumstances and implementation decisions that relate to problematic AI-generated solutions. Table~\ref{tab:rq2_quant} shows that respondents identified ambiguous requirements (59.1\%), complex business logic (53.3\%), and large-scale systems (43.1\%) as situations in which AI struggled to reflect requirements. They also reported overreliance on AI-generated solutions (58.4\%), reduced knowledge sharing (54.7\%), and inconsistent development practices (51.1\%).

Qualitative accounts illustrated how insufficient project context influenced implementation decisions. In Q34 (Table~\ref{tab:q34-themes}), 23 respondents (21.1\%) described Context Misalignment. S14 explained, \emph{``If you provide a direct prompt explaining the requirements only without any guidelines on system design, it will find a quick temporary solution with high coupling and lack of scalability.''} P1 described a rushed project in which developers provided insufficient documentation and delegated implementation decisions to AI, resulting in unnecessary validation checks, functions, and repetitive API calls. P15 reported that AI modified intentional business-specific behavior, while P14 described how AI implemented subscription tiers without clarifying requirements: \emph{``it built various subscription layers on its own, and did the whole implementation on its own, without asking what features there would be, or what kind of things to keep, or what to keep in the free tier.''}

Development pressure and limited review capacity also influenced practitioners' decisions. In Q26 (Table~\ref{tab:rq2_quant}), respondents reported unrealistic productivity expectations (48.9\%) and reduced code review quality (47.4\%). S12 described pushing AI-generated code without review because of deadline pressure. P11 reported generating approximately 65,000 lines of code in five days across numerous files, explaining, \emph{``Now, reviewing that code has been going on for two weeks, and I still haven't been able to visit all of the files.''} This experience illustrated how rapid generation could exceed developers' review capacity.

Respondents also identified architectural and design problems. In Q14, they reported inconsistent architecture patterns (53.3\%), maintainability issues (46.7\%), and tight coupling (36.5\%). In Q20 (Table~\ref{tab:rq2_quant}), they reported temporary workarounds (49.6\%), overly long functions or classes (40.9\%), weak abstractions or design patterns (38.0\%), and poor reusability (38.0\%). These accounts connected unclear requirements, insufficient architectural guidance, deadline pressure, and limited review with implementation problems that may increase maintenance effort.

\summarybox{Respondents identified ambiguous requirements (59.1\%) and overreliance on AI-generated solutions (58.4\%) as prominent concerns, alongside inconsistent architectural patterns (53.3\%) and temporary workarounds (49.6\%). Qualitative accounts linked insufficient project guidance, requirements misalignment, deadline pressure, and large volumes of generated code to implementation and review difficulties.}

\begin{table}[tb]
\centering
\scriptsize
\renewcommand{\arraystretch}{1.1}
\caption{Top five reported development conditions and implementation issues in AI-assisted development ($N=137$; multiple selections allowed).}
\label{tab:rq2_quant}
\resizebox{\columnwidth}{!}{%
\begin{tabular}{|p{4.2cm}|c|l|c|} \hline
\multicolumn{2}{|l|}{\textbf{Q14: Architecture-related issues}} & \multicolumn{2}{l|}{\textbf{Q20: Design-related issues}} \\
\hline
\textbf{Response} & \textbf{\%} & \textbf{Response} & \textbf{\%} \\
\hline
Inconsistent architecture patterns & 53.3\% & Temporary workaround implementations & 49.6\% \\
Maintainability & 46.7\% & Overly long functions / classes & 40.9\% \\
Tight coupling between components & 36.5\% & Weak abstractions or design patterns & 38.0\% \\
Difficult system integration & 29.9\% & Poor reusability & 38.0\% \\
Poor scalability & 29.2\% & Poor separation of concerns & 27.7\% \\
\hline
\multicolumn{2}{|l|}{\textbf{Q26: Management-related issues}} & \multicolumn{2}{l|}{\textbf{Q30: Requirements-related situations}} \\
\hline
\textbf{Response} & \textbf{\%} & \textbf{Response} & \textbf{\%} \\
\hline
Over-reliance on AI-generated solutions & 58.4\% & Ambiguous requirements & 59.1\% \\
Reduced knowledge sharing within teams & 54.7\% & Complex business logic & 53.3\% \\
Inconsistent development practices across team members & 51.1\% & Large-scale systems & 43.1\% \\
Unrealistic productivity expectations & 48.9\% & Client-specific customization & 40.1\% \\
Reduced code review quality & 47.4\% & Multi-step workflows & 36.5\% \\
\hline
\end{tabular}
}
\end{table}

%% file: results/rq3.tex
To answer RQ3, we examine technical problems that practitioners encountered during AI-assisted development and their maintenance consequences. Table~\ref{tab:rq3_quant} summarizes the code, configuration, documentation, and infrastructure problems. In Q16 (Table~\ref{tab:rq3_quant}), respondents most frequently reported overly complex implementations (68.6\%), hard-to-maintain code (46.0\%), and duplicated code (43.1\%). Configuration issues included unnecessary dependencies (42.3\%), deprecated libraries or frameworks (37.2\%), and version conflicts (35.8\%). Other concerns included inaccurate or hallucinated documentation (40.9\%), generic or non-useful documentation (33.6\%), security issues in infrastructure setup (35.0\%), and scalability or reliability problems (26.3\%).

Open-ended responses highlighted debugging and maintenance challenges. In Q34 (Table~\ref{tab:q34-themes}), 26 respondents reported Maintainability issues, 15 reported Code Duplication, and 7 reported Debugging Difficulties. S1 described duplicated state-management logic that caused intermittent synchronization bugs over several weeks. S3 reported similar issues in a six-stage approval system: \emph{``If I need to change a value then I need to change 6 different places. The basic version was developed by AI which was a technical debt for me.''} S90 described an AI-generated module with non-standard logic that the team spent two days debugging before rewriting it from scratch. P12 similarly noted that AI sometimes duplicated logic across files, requiring explicit instructions to create reusable functions.

Rework emerged as a prominent consequence, which 33 respondents reported in Q34. S64 described inconsistent architectural patterns and excessive coupling in AI-generated service and database layers that required architectural refactoring. S17 reported incorrect dependency versions in an AI-generated configuration file that disrupted development and required manual correction. Similarly, 80\% (16/20) of interviewees reported corrective work ranging from architectural restructuring to removing unrequested changes.

\begin{table}[tb]
\centering
\scriptsize
\renewcommand{\arraystretch}{1.1}
\caption{Top five reported problems in AI-assisted development across code, configuration, documentation, and infrastructure ($N=137$; multiple selections allowed).}
\label{tab:rq3_quant}
\resizebox{\columnwidth}{!}
{%
\begin{tabular}{|l|c|l|c|} \hline
\multicolumn{2}{|l|}{\textbf{Q16: Code-related problems}} &
\multicolumn{2}{l|}{\textbf{Q18: Configuration-related problems}} \\
\hline
\textbf{Response} & \textbf{\%} &
\textbf{Response} & \textbf{\%} \\
\hline
Overly complex implementations & 68.6\% &
Unnecessary dependencies & 42.3\% \\
Hard-to-maintain code & 46.0\% &
Deprecated libraries / frameworks & 37.2\% \\
Duplicated code & 43.1\% &
Version conflicts & 35.8\% \\
Excessive unused code & 31.4\% &
Compatibility issues across environments & 29.2\% \\
Poor readability & 29.9\% &
Incorrect environment configurations & 28.5\% \\
\hline
\multicolumn{2}{|l|}{\textbf{Q22: Documentation-related problems}} &
\multicolumn{2}{l|}{\textbf{Q24: Infrastructure-related problems}} \\
\hline
\textbf{Response} & \textbf{\%} &
\textbf{Response} & \textbf{\%} \\
\hline
Hallucinated or inaccurate information & 40.9\% &
Security issues in infrastructure setup & 35.0\% \\
Generic / non-useful documentation & 33.6\% &
Scalability or reliability problems & 26.3\% \\
Outdated explanations & 24.8\% &
Poor monitoring / logging setup & 24.1\% \\
Missing documentation & 22.6\% &
Cloud resource misconfiguration & 21.9\% \\
Inconsistent documentation & 20.4\% &
Inefficient deployment scripts & 20.4\% \\
\hline
\end{tabular}%
}
\end{table}

Three interviewees reported performance issues, and the unrequested changes that RQ1 reports also surfaced during validation. P2 described an AI-assisted implementation that slowed after deployment to another environment: \emph{``But it turns out that when I take it to another server--say it was on a training server and I move it to production--it becomes a bit slow there.''} P4 described an unexplained modification during a Bootstrap migration: \emph{``Then during testing and review we saw that it made this unnecessary change, and we couldn't really explain it.''} 
Interviewees reported increased review burdens (35\%, 7/20), weakened code ownership (10\%, 2/20), and production risks (20\%, 4/20). P6 raised concerns about code ownership under heavy AI use, which \emph{``basically belongs to no one anymore.''} P9 highlighted the risks of undetected issues reaching production: \emph{``when that comes back as a production issue, that becomes a much bigger problem. At that point, we're the ones to blame.''}

\summarybox{
Respondents most frequently identified overly complex implementations (68.6\%), hard-to-maintain code (46.0\%), and duplicated code (43.1\%), alongside configuration, documentation, and infrastructure problems. Qualitative accounts revealed debugging, refactoring, and deployment issues, while interviews highlighted review burdens and weakened code ownership. These consequences indicated potential, but not necessarily persistent, technical debt.
}

%% file: results/rq4.tex
To answer RQ4, we examine how practitioners validate AI-generated changes and manage potential technical debt risks. Table~\ref{tab:rq4_quant} shows that 89.8\% of respondents reported always or often verifying AI-generated code before use. Manual code review was the most common validation method (78.1\%), ahead of automated testing (46.0\%) and documentation checks (45.3\%). Respondents also reported writing additional tests (35.0\%), conducting peer reviews (32.8\%), and manually rewriting generated code (27.7\%). However, 10.2\% selected ``rarely validate'' in Q32 (data not present in Table~\ref{tab:rq4_quant}), indicating differences in verification practices. These responses reflect reported practices rather than their effectiveness in preventing technical debt.

Open-ended responses provided further evidence of practices for managing technical debt risks. In Q33 (Table~\ref{tab:q33-themes}), of 118 respondents, 46 (39.0\%) described Refactoring, 36 (30.5\%) described Code Quality \& Maintainability improvements, and 31 (26.3\%) described Test Generation. In Q35 (Table~\ref{tab:q35-themes}), of 112 respondents, suggested improvements involved Architectural Awareness (30, 26.8\%), Risk Identification (18, 16.1\%), Testing Support (17, 15.2\%), Automated Code Validation (12, 10.7\%), and Human Oversight (9, 8.0\%). S22 advocated incremental implementation: \emph{``instead of generating full feature at once, just write one step, and review it with the developer.''}

Interview participants described practices for controlling AI-generated changes before implementation. Fifteen interviewees (75.0\%) provided detailed requirements, instructions, or project documentation, while 11 (55.0\%) reviewed or refined AI-generated plans and six (30.0\%) divided tasks into smaller steps. P12 described repeatedly reviewing plans to identify missing requirements and incorrect assumptions, explaining, \emph{``Only after the plan is finalized do I assign the actual implementation work.''} Nine interviewees (45.0\%) also restricted the files or scope that AI could modify.

Participants also emphasized reviewing generated implementations. P16 explained, \emph{``Since my prompts are small, or are for specific tasks, the AI-generated code is also very small. So it's not much of a hassle for me to review the code.''} P16 manually inspected changes and tested their functionality, while three interviewees (15.0\%) explicitly limited change size to facilitate review. Across the interviews, 14 participants (70.0\%) tested generated code, nine (45.0\%) checked architectural consistency, nine (45.0\%) used another AI tool for cross-checking, six (30.0\%) consulted documentation or conducted additional research, four (20.0\%) examined edge cases, and five (25.0\%) inspected repository changes or pull requests.

When AI introduced unnecessary changes (I-Q3), 12 interviewees (60.0\%) reverted, removed, or corrected them, while six (30.0\%) tightened their prompts or scope instructions. P3 distinguished useful additions, such as overlooked edge cases, from unnecessary modifications and instructed AI to revert the latter.

\summarybox{
Most respondents (89.8\%) reported always or often verifying AI-generated code, with greater reliance on manual review (78.1\%) than automated testing (46.0\%). Interviewees described reviewing plans before delegation (11 of 20), testing generated code (14 of 20), and correcting unwanted changes (12 of 20). Our study does not measure the effectiveness of these practices in reducing technical debt.
}

\begin{table}[tb]
\centering
\scriptsize
\caption{Reported verification and validation practices for AI-generated code ($N=137$; multiple selections allowed for Q32).}
\label{tab:rq4_quant}
% \resizebox{\columnwidth}{!}
{%
\begin{tabular}{|l|c|l|c|} \hline
\multicolumn{2}{|l|}{\textbf{Q11: Verification frequency}} &
\multicolumn{2}{l|}{\textbf{Q32: Validation methods (Top 5)}} \\
\hline
\textbf{Response} & \textbf{\%} &
\textbf{Response} & \textbf{\%} \\
\hline
Always & 55.5\% & Manual code review & 78.1\% \\
Often & 34.3\% & Running automated tests & 46.0\% \\
Sometimes & 6.6\% & Cross-checking documentation & 45.3\% \\
Rarely & 2.9\% & Writing additional tests & 35.0\% \\
Never & 0.7\% & Peer review & 32.8\% \\
\hline
\end{tabular}%
}
\end{table}

%% file: results/rq5.tex
To answer RQ5, we examine how practitioners use AI for learning, delegate development tasks, and make technical decisions. Table~\ref{tab:rq5-developer-practices} shows that 93.4\% of respondents used AI multiple times a day, while 96.4\% considered it helpful or very helpful. Common uses included code generation (86.9\%), debugging (84.7\%), code explanation (78.1\%), and documentation (77.4\%). Most respondents reported significant (64.2\%) or some (32.1\%) learning benefits. However, only 44.5\% mostly or completely trusted unmodified AI-generated code. Respondents also identified over-dependence on AI tools (73.7\%), reduced understanding of generated code (64.2\%), and skill degradation (64.2\%) as people-related challenges.

Open-ended responses reflected both benefits and risks. In Q33 (Table~\ref{tab:q33-themes}), respondents highlighted gains in Productivity (37, 31.4\%), Code Comprehension (25, 21.2\%), and Debugging (18, 15.3\%). In Q34 (Table~\ref{tab:q34-themes}), S132 reported declining hands-on problem-solving and skill development as team members increasingly relied on AI. In Q35 (Table~\ref{tab:q35-themes}), respondents requested Explainability (13, 11.6\%) and Developer Skill Development (11, 9.8\%). S84, for example, wanted AI to explain code in simple language and identify potential issues to support learning.

Interview participants described how AI influenced their development decisions. P11 explained, \emph{``Before, I used to think about how to solve the problem. Now, the first thing that comes to my mind is how I can explain the problem to AI in a better way.''} This account illustrated a shift toward preparing instructions and context for AI. When deciding what to delegate (I-Q2), four interviewees prioritized time savings, while four delegated coding tasks but retained control over business logic or major decisions.

Participants also described learning strategies for unfamiliar tasks (I-Q6). Nineteen interviewees (95.0\%) reported developing their understanding before delegating unfamiliar work, including 15 (75.0\%) who asked AI for explanations and six (30.0\%) who consulted documentation or other references. P3 explained, \emph{``if I have the concept clear beforehand, then whatever kind of issue comes up, I'll be able to fix it easily.''} P18 similarly described evaluating alternative approaches before implementation and emphasized, \emph{``I keep control of it. I do the R\&D on it myself first, build a small project to try it out.''} Three participants (15.0\%) initially delegated unfamiliar tasks directly but later adopted a learning-first approach, while P18 allowed greater AI autonomy for simple applications. In contrast, P9 used AI to build a project in an unfamiliar technology stack and could verify its behavior but not its underlying code.

P13 described increasing dependence on AI, acknowledging, \emph{``I mean, my level of human thinking has dropped a lot. Second, the dependency has become much greater.''} P13 also reported occasionally pushing AI-generated code without adequate review. This account contrasted with the learning-first approach that P3 described and illustrated concerns about declining independent problem-solving. Although survey respondents identified over-dependence (73.7\%) and skill degradation (64.2\%) as risks, their responses do not establish measurable changes in developer ability.

\summarybox{
Most respondents reported significant learning benefits (64.2\%), alongside concerns about over-dependence (73.7\%) and reduced understanding of generated code (64.2\%). Nineteen of 20 interviewees reported developing their understanding before delegating unfamiliar tasks, while interview accounts revealed contrasting experiences with AI dependence and independent problem-solving.
}

\begin{table}[t]
\centering
\scriptsize
\caption{Reported developer practices, learning, and decision-making in AI-assisted development ($N=137$; multiple selections allowed for Q8 and Q28).}
\label{tab:rq5-developer-practices}
\resizebox{\columnwidth}{!}
{%
\begin{tabular}{|l|c|l|c|} \hline
\textbf{Response} & \textbf{\%} &
\textbf{Response} & \textbf{\%} \\
\hline
\multicolumn{2}{|l|}{\textbf{Q10: Trust in unmodified code}} &
\multicolumn{2}{l|}{\textbf{Q7: AI-tool usage frequency}} \\
\hline
Completely trust & 2.9\% & Multiple times a day & 93.4\% \\
Mostly trust & 41.6\% & Once a day & 1.5\% \\
Neutral & 29.9\% & Several times a week & 2.2\% \\
Slightly trust & 19.0\% & Several times a month & 1.5\% \\
No trust at all & 6.6\% & Rarely & 1.5\% \\
\hline
\multicolumn{2}{|l|}{\textbf{Q12: Learning new concepts}} &
\multicolumn{2}{l|}{\textbf{Q8: Purposes of AI-tool use (Top 5)}} \\
\hline
Significantly & 64.2\% & Code generation & 86.9\% \\
Somewhat & 32.1\% & Bug fixing / debugging & 84.7\% \\
Rarely & 3.6\% & Explaining code & 78.1\% \\
Never & 0.0\% & Writing documentation & 77.4\% \\
 & & Brainstorming ideas & 72.3\% \\
\hline
\multicolumn{2}{|l|}{\textbf{Q9: Perceived helpfulness}} &
\multicolumn{2}{l|}{\textbf{Q28: People-related challenges (Top 5)}} \\
\hline
Very helpful & 66.4\% & Over-dependence on AI tools & 73.7\% \\
Helpful & 29.9\% & Reduced understanding of generated code & 64.2\% \\
Neutral & 1.5\% & Skill degradation / problem-solving reduction & 64.2\% \\
Slightly helpful & 2.2\% & Reduced collaboration among team members & 59.9\% \\
Not helpful & 0.0\% & Communication gaps during development & 48.9\% \\
\hline
\end{tabular}%
}
\end{table}

%% file: discussion.tex
\section{Discussion and Implications}

\subsection{Technical Debt Risks Beyond Code Quality}

Technical debt risks in AI-assisted development extend beyond code quality to encompass broader socio-technical concerns. The prominence of People Debt, Management Debt, and Architecture Debt (RQ1) suggests that a code-centric perspective alone is insufficient to capture these risks, aligning with research on the human and organizational dimensions of technical debt~\cite{caglayan2024revisiting, Ahmad2022PeopleDebt}. Prior research also emphasized the role of human judgment in technical debt management and introduced concepts such as cognitive and intent debt~\cite{avgeriou2026technical, storey2026technical}. The contrasting experiences reported in Q33 and Q34 indicate that perceived improvements in refactoring, maintainability, and testing can coexist with rework, context misalignment, and maintenance difficulties, making the long-term implications of AI-assisted development difficult to assess. Such risks may not be immediately apparent in generated code but can arise through implementation decisions, coordination challenges, and subsequent maintenance.

Interview accounts suggest that insufficient project context and unclear requirements can lead to implementations that conflict with existing architecture or business constraints. Prior research reported similar concerns~\cite{sergeyuk2025using, mu2024clarifygpt}. Interviewees also noted that rapid code generation can increase the effort needed to understand and validate changes. This aligns with prior studies reporting faster task completion~\cite{peng2023impact} but greater review and rework effort~\cite{xu2025aiassisted}. Consequently, AI-assisted development may shift effort from implementation to review, correction, and maintenance, potentially offsetting productivity gains when developers poorly understand generated changes or when those changes conflict with project requirements.

\subsection{AI Delegation, Developer Judgment, and Maintenance}

The perceived learning benefits and concerns about AI dependence (RQ5) suggest that AI delegation can both support developers and reduce their engagement with implementation details. The contrasting practices of understanding unfamiliar concepts before delegation (P3) and verifying application behavior without fully understanding the code (P9) illustrate how developers may retain different levels of involvement in AI-assisted tasks. Prior research similarly identified changes in developers' decision-making, cognitive engagement, and collaboration~\cite{tavantzis2026displaces}, as well as difficulties understanding and modifying generated solutions~\cite{vaithilingam2022expectation, mujahid2026todo}. These concerns suggest that the consequences of AI delegation depend not only on task completion but also on whether developers retain sufficient understanding to evaluate and modify the resulting implementation.

Manual review, testing, and implementation planning (RQ4) remain important validation practices, consistent with prior research~\cite{tabarsi2026reshaping, tang2024validating}. However, difficulties understanding AI's implementation decisions (RQ1) and maintaining unfamiliar generated code (RQ3) suggest that functional verification alone may not preserve sufficient knowledge for future maintenance. Developers should therefore emphasize understanding implementation rationale alongside functional verification to preserve developer judgment, code ownership, and accountability~\cite{avgeriou2026technical}.

\subsection{Implications for Development Practices and Tools}

Ambiguous requirements and insufficient project context suggest that developers should clarify implementation intent and establish clearer implementation boundaries before delegating tasks to AI assistants. Coding assistants could retain project conventions, ask targeted questions, and present implementation plans for review~\cite{mu2024clarifygpt}. To address unrequested changes and excessive review effort, tools could distinguish requested changes from optional improvements and support smaller, incremental modifications.

Teams should also strengthen verification and preserve implementation rationale. Beyond checking functional correctness, developers should review generated code for architectural consistency, maintainability, and alignment with project requirements~\cite{tang2024validating, liu2024refining}. To address difficulties recovering implementation rationale, coding assistants could record assumptions, design choices, and explanations alongside code changes to support future maintenance. These practices could address decision-traceability challenges and support human-driven technical debt management~\cite{carvalho2026traceability, avgeriou2026technical}.

\subsection{Implications for Researchers and Future Studies}

The contrasting experiences with AI-assisted development warrant further investigation into the conditions under which AI reduces maintenance effort or introduces additional technical debt. Prior repository studies investigated technical debt indicators in AI-authored code~\cite{liu2026debt, niemeyer2026genai}. Future longitudinal studies could trace developers' intentions, AI-generated modifications, code review decisions, and subsequent corrective work to distinguish temporary implementation problems from persistent technical debt. Researchers should also examine how AI delegation affects developer understanding, decision ownership, and knowledge transfer within teams, extending socio-technical perspectives on technical debt~\cite{Ahmad2022PeopleDebt, storey2026technical, tavantzis2026displaces}.

Future research should also empirically evaluate practices and tool capabilities intended to mitigate technical debt risks. Controlled studies could assess whether architectural awareness, requirements clarification, incremental implementation, and automated validation reduce unnecessary changes, architectural inconsistencies, and rework. Prior studies reported contrasting maintenance outcomes~\cite{xu2025aiassisted, he2026speed, borg2026echoes}. Comparing different delegation and verification strategies across project sizes, developer experience levels, and AI-tool configurations could help explain these differences and identify conditions that influence long-term maintainability.

%% file: conclusion.tex
\section{Threats to Validity}

\noindent
\textbf{Construct Validity.} We adopt ten technical debt dimensions from Caglayan et al.~\cite{caglayan2024revisiting} and develop survey items using established indicators and concerns in AI-assisted development. However, these dimensions and predefined response options may not capture all emerging risks and may influence participants' selections. Participants may also interpret implementation problems as technical debt even when developers correct them before they create lasting maintenance costs. We address these concerns through open-ended questions (Q33-Q35) and follow-up interviews that capture experiences beyond predefined categories.

\smallskip
\noindent
\textbf{Internal Validity.} Our study relies on practitioners' self-reported experiences, which may involve recall bias, subjective interpretations, and selective reporting. Participants may attribute implementation problems to AI even when existing codebase conditions, incomplete requirements, time pressure, or developer decisions contribute to the outcomes. We use follow-up interviews to examine these circumstances, although we cannot independently verify reported incidents or establish causal relationships. Researcher interpretation may also influence our qualitative findings. To improve consistency, two annotators rigorously examine open-ended responses, compare their codes, and resolve disagreements through discussion. Nevertheless, different researchers may interpret ambiguous responses differently.

\smallskip
\noindent
\textbf{External Validity.} Our study includes 137 survey respondents and 20 interview participants from the software industry in Bangladesh. We recruit participants through professional and social media networks, which introduces self-selection bias and may overrepresent practitioners with strong opinions or greater experience using AI coding assistants. Although participants represent different roles and experience levels, organizational practices, project characteristics, and AI-tool usage may differ across regions and environments, thus, we characterize participants' perceptions and experiences.

\section{Conclusion}
Our findings reveal technical debt risks across all ten dimensions, with the highest agreement for People Debt, Management Debt, and Architecture Debt. Drawing on 137 survey responses and 20 interviews, we highlight the socio-technical nature of these risks, the importance of developer judgment, understanding, and code ownership, and the need for mitigation practices. Practitioners identify ambiguous requirements, insufficient project context, AI overreliance, and review difficulties as prominent concerns. These findings emphasize that technical debt risks arise not only from generated code but also from the decisions and practices surrounding its development.

We also identify a tension between developers' intended improvements through refactoring, testing, and code-quality practices and their reported experiences of rework, context misalignment, and maintenance difficulties. As AI assistants assume greater implementation responsibility, managing technical debt risks requires more than verifying generated code. Developers and teams need to clarify requirements, preserve project context and implementation rationale, control generated changes, and retain sufficient understanding to maintain the resulting software.